\pdfoutput=1

\documentclass[11pt]{article}

\usepackage{acl}

\usepackage{times}
\usepackage{latexsym}
\usepackage{amsfonts}

\usepackage[T1]{fontenc}

\usepackage[utf8]{inputenc}

\usepackage{microtype}

\usepackage{graphicx}
\usepackage{color,soul}
\usepackage{amsmath}

\title{AB/BA analysis: A framework for estimating keyword spotting recall improvement while maintaining audio privacy}

\author{
Raphael Petegrosso, Vasistakrishna Baderdinni, Thibaud Senechal\footnotemark[1], Benjamin L. Bullough\footnotemark[1] \\
  Amazon / USA\\
\{petegrr,vbaderdi,thibauds,bullough\}@amazon.com

 }

\usepackage{fancyhdr} 
\begin{document}
\maketitle

\def\thefootnote{*}\footnotetext{Corresponding authors}

\begin{abstract}
Evaluation of keyword spotting (KWS) systems that detect keywords in speech is a challenging task under realistic privacy constraints. The KWS is designed to only collect data when the keyword is present, limiting the availability of hard samples that may contain false negatives, and preventing direct estimation of model recall from production data. Alternatively, complementary data collected from other sources may not be fully representative of the real application. In this work, we propose an evaluation technique which we call AB/BA analysis. Our framework evaluates a candidate KWS model $B$ against a baseline model $A$, using cross-dataset offline decoding for relative recall estimation, without requiring negative examples. Moreover, we propose a formulation with assumptions that allow estimation of relative false positive rate between models with low variance even when the number of false positives is small. Finally, we propose to leverage machine-generated soft labels, in a technique we call Semi-Supervised AB/BA analysis, that improves the analysis time, privacy, and cost. Experiments with both simulation and real data show that AB/BA analysis is successful at measuring recall improvement in conjunction with the trade-off in relative false positive rate. 
\end{abstract}

\section{Introduction}

Keyword spotting (KWS) is the task of identifying if one of a set of keywords, also called wakewords, is present in a speech segment. It is the gatekeeper component that enables hands-free interaction with many smart assistants using voice-enabled smart devices, such as Amazon Echo, Google Home, and Apple HomePod. Extensive work has been done to develop and improve KWS performance, including improvements in architecture \citep{DBLP:conf/icassp/ChenPH14, DBLP:conf/interspeech/SunSGNRPSMV17, DBLP:conf/interspeech/ShanZWX18, DBLP:conf/icassp/WuPSGTVHM18, DBLP:conf/interspeech/GaoSKCSZV20}, training efficiency \citep{DBLP:conf/interspeech/TuckerWSPFV16, DBLP:journals/corr/abs-1808-00563}, as well as audio front-end (AFE) algorithms \citep{DBLP:conf/eusipco/ChhetriHKCMLZ18}. 

With many lines of research aiming to improve the quality of the KWS, there is also growing interest in techniques to measure if such new technologies are able to improve the customer experience, but less research attention has been given to this evaluation topic. One challenge is that KWS systems are designed to maximize user privacy by only collecting data when the keyword is identified. As a result, evaluations done using this biased dataset do not allow direct measurement of gains in recall metrics to determine if a new model is better than a baseline. Alternatively, evaluations can also make use of datasets not collected by the KWS, such as media recordings, background noise, and environmental sounds. However, these datasets may not be fully representative of the user experience from the evaluation point of view.

Prior research, such as from \citep{DBLP:conf/interspeech/GaoSKCSZV20, DBLP:conf/interspeech/SainathP15}, has made use of datasets with positive and negative labels in order to evaluate, respectively, gain invariant KWS models and CNN for small-footprint KWS. However, the negative data essentially consists of either negative labels obtained from data accepted by the previous models, or datasets composed of just background noise and noise from environment. In a different application context, \citep{DBLP:conf/sdm/MillerVKZ18} estimates the recall and the derivative of the precision with respect to the recall by modeling the unseen data distribution according to underlying assumptions. This distribution, however, is not clearly defined in the context of KWS when a variety of sounds, noisy conditions, and reverberation can occur. 

In this paper, we present a new evaluation framework called AB/BA analysis. To the best of our knowledge, our work is the first to explore the problem of estimating recall improvement when only accepted data is available, such as in a KWS system with rigorous privacy settings, in conjunction with the trade-off with false positive rate, without underlying assumptions on the data distribution. We show that the AB/BA analysis is a cross-model offline evaluation framework. In this framework, data is collected from two KWS models, say a baseline, $A$ , and a candidate, $B$. By running offline model $A$ through data collected by model $B$, and model $B$ through data collected by model $A$, we are able to calculate relative metrics without the need for data not seen by individual models. We also present assumptions that can be applied to the relative metrics calculation that result in a lower variance estimator, even if the number of false positive is small. We additionally describe a Semi-Supervised formulation of AB/BA analysis which provides improvements in the analysis time, cost, and data privacy by utilizing soft machine-generated labels instead of human annotations from the models being evaluated. We present experiments using simulation data, real data, and also experiments comparing the performance of AB/BA and Semi-Supervised AB/BA.

\section{Methods}\label{sec:methods}

\begin{figure*}[h]
\includegraphics[width=\textwidth]{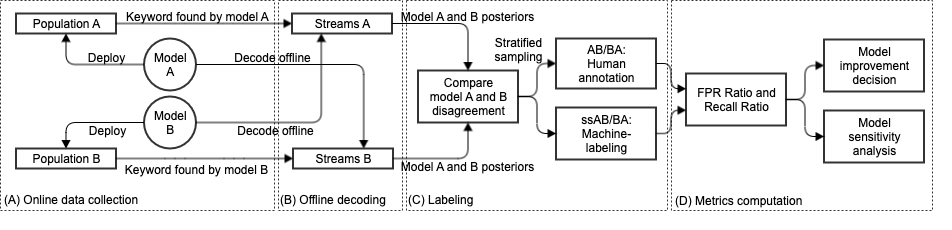}
\caption{\textbf{The four main components of the AB/BA analysis}}\label{fig:abba_diagram}
\end{figure*}

In this section, we describe the proposed method for estimation of recall improvement. Basic metrics concepts can be found in Appendix \ref{sec:metrics}.

\subsection{AB/BA Analysis}\label{sec:abba}

The AB/BA analysis framework is composed of four main steps, as depicted in Figure \ref{fig:abba_diagram}: Online data collection, Offline decoding, Labeling, and Metrics computation.

In order to collect data, given two KWS models, say $A$ and $B$, the models are deployed simultaneously to two populations of users, also called $A$ and $B$, as shown in Figure \ref{fig:abba_diagram} (A). The percentage of users in each model is usually based on the presumed risk of deploying each model, as well as the statistical significance desired for the metrics computed, as shown later. It is important, though, that models are deployed simultaneously, with random assignment, similar to a conventional A/B-Test. Notice that the data collected by the models will only contain samples where the keyword is detected in order to preserve user privacy. 

The collected data is then used for offline decoding, in which models are run offline on the data collected online. As shown in Figure \ref{fig:abba_diagram} (B), the data collected by model $A$ is used by model $B$ for keyword spotting, and model $B$ is run on the data collected by the model $A$. A detailed review on keyword spotting can be found in \cite{DBLP:journals/corr/abs-2111-10592}. Running a keyword spotter on an utterance $i$ will produce a score $s_i$ representing how likely it is to have detected the keyword. 

The data that is collected and decoded offline by the KWS systems will also require labeling in order to be used for metrics estimation. This step, shown in Figure \ref{fig:abba_diagram} (C),  is usually done by either human annotation, or machine-generated. The process to use machine-generated labels is described in more detail in Section \ref{sec:ssabba}.

Human annotation is an expensive and time consuming process, in addition to being susceptible to error. Therefore, it is desirable to carefully select which utterances to annotate. Stratified sampling can be used to provide more annotations on models disagreement to reduce the need for human annotations (details in Appendix \ref{appendix:stratified}). The labeled data is then used to compute metrics that represent the relative improvement of model $B$ with respect of a baseline model $A$, as shown in Figure \ref{fig:abba_diagram} (D). 

AB/BA analysis utilizes two relative metrics: False Positive Rate Ratio and Recall Ratio. The Recall Ratio (rRecall) is a relative metric used for comparing two models to determine which model yields better recall. Given two KWS models $A$ and $B$ with datasets containing utterances collected by the same models online, and labeled as $L \in \{0,1\}$, using the Bayes' theorem, the rRecall can be defined as:

\begin{equation}\label{eq:recall_ratio_form}
\begin{split}
rRecall &= \frac{P(B=1|L=1)}{P(A=1|L=1)} \\
&= \frac{P(B=1|A=1,L=1)}{P(A=1|B=1,L=1)},
\end{split}
\end{equation}
where $P(B=1|A=1,L=1)$ indicates the probability model $B$ found the keyword when run offline ($B=1$) on true positives from model $A$ used online ($A=1,L=1$).

Therefore, a key aspect of AB/BA analysis is that the rRecall can be computed using terms $P(B=1|A=1,L=1)$ and $P(B=1|A=1,L=0)$ that are directly observable.

Concretely, we can compute rRecall using the following quantities:

\begin{equation}\label{eq:recall_ratio}
rRecall = \frac{NTP_{BA\_on\_A}}{NPos_{A}} * \frac{NPos_{B}}{NTP_{AB\_on\_B}},
\end{equation}
where $NTP_{AB\_on\_B}$ is the number of TPs of model $A$ and $B$ on data collected by model $B$ and $NPos_{B}$ is the number of positive labels on the data collected by $B$. 


Similarly, the FPR Ratio (rFPR) between two models $A$ and $B$ can be calculated as:

\begin{equation}\label{eq:fa_ratio}
rFPR = \frac{NFP_{BA\_on\_A}}{NNeg_{A}} *\frac{NNeg_{B}}{NFP_{AB\_on\_B}},
\end{equation}
where $NFP_{AB\_on\_B}$ is the number of FPs of model $A$ and $B$ on data collected by model $B$ and $NNeg_{B}$ is the number of negative labels on the data collected by $B$. 


Notice that the number of FPs can be small, leading to large variance in the rFPR estimation. Therefore, assuming that keywords and confusing sounds (those that induce FPs) generated by population $A$ and population $B$ are randomly drawn from the same distribution, we propose to assume that the ratio of TPs and FPs in the streams accepted by both models is the same, represented as:

\begin{equation}
\frac{NFP_{BA\_on\_A}}{NTP_{BA\_on\_A}} \approx \frac{NFP_{AB\_on\_B}}{NTP_{AB\_on\_B}}
\end{equation}

Then, by introducing the following variables:

\begin{equation}
\begin{split}
NTP_{AB} = NTP_{BA\_on\_A} + NTP_{AB\_on\_B}\\
NFP_{AB} = NFP_{BA\_on\_A} + NFP_{AB\_on\_B}\\
\alpha=\frac{NTP_{BA\_on\_A} + NFP_{BA\_on\_A}}{NTP_{AB} + NFP_{AB}}\\
\beta=\frac{NTP_{AB\_on\_B} + NFP_{AB\_on\_B}}{NTP_{AB} + NFP_{AB}}
\end{split}
\end{equation}

We can find rFPR and rRecall using:

\begin{equation}\label{eq:approx_method}
\begin{split}
rFPR = \frac{\alpha(NFP_B + \beta NFP_{AB})}{\beta(NFP_A + \alpha NFP_{AB})}\\
rRecall = \frac{\alpha(Nmiss_A + \beta NTP_{AB})}{\beta(Nmiss_B + \alpha NTP_{AB})},
\end{split}
\end{equation}
where $Nmiss_A$ is the number of TPs accepted by model $B$ but not accepted by $A$. With this estimator, uncertainties on $NFP_{AB}$ have less impact on the rFPR uncertainty. This is shown in the simulation on Section \ref{sec:abba_sim}.

\subsection{Semi-Supervised AB/BA Analysis} \label{sec:ssabba}

Semi-Supervised AB/BA (ssAB/BA) analysis is a technique to estimate rFPR and rRecall metrics, while avoiding the need to label utterances by human annotation, which are instead estimated in a semi-supervised way. Because of that, the technique has lower cost and is faster to run than AB/BA analysis. The process also improves audio privacy, since no audio is listened to by annotators. However, since it relies on a machine-labeling process, the technique is more susceptible to errors due to bias.

There are several lines of research on Semi-Supervised Learning models, such as Teacher models \citep{li2017large, tarvainen2017mean}, which provide posterior probabilities  as soft labels in order to train Student models.

Assuming that we have a Label Machine $\mathcal{M}$, we apply this machine on utterance $i$ to produce a score $m_i$. However, if $m_i$ produces a soft label $m_i$, a mapping function $\phi_{\mathcal{M}}(m_i) = p_i$ can be used to convert the machine-generated score $m_i$ from machine $\mathcal{M}$ to a probability of true accept $p_i$. Appendix \ref{appendix:calibration} illustrates this process using a polynomial mapping.

When only soft labels are available, representing a probability of true label, we can apply the Bayes' theorem on Equation $\eqref{eq:recall_ratio_form}$ to calculate rFPR using:

\begin{equation}
rFPR = \frac{\frac{P(L=0|B=1,A=1)*P(B=1|A=1)}{P(L=0|A=1)}}{\frac{P(L=0|A=1,B=1)*P(A=1|B=1)}{P(L=0|B=1)}},
\end{equation}
which results in:

\begin{equation}\label{eq:fa_ratio_ssl}
\begin{split}
& rFPR = \frac{\sum_{i=0}^{N_A}p(L_i=0 | A_i=1, B_i=1)}{\sum_{i=0}^{N_A}p(L_i=0 | A_i=1)} * \\
& \frac{\sum_{i=0}^{N_B}p(L_i=0 | B_i=1)}{\sum_{i=0}^{N_B}p(L_i=0 | B_i=1, A_i=1)},
\end{split}
\end{equation}

Equations $\eqref{eq:fa_ratio}$ and $\eqref{eq:fa_ratio_ssl}$ are, therefore, equivalent if $p(L_i=0)$ is a hard ground-truth label (either 0 or 1). 

Similarly, in Semi-Supervised AB/BA the rRecall can be written as:

\begin{equation}\label{eq:recall_ratio_ssl}
\begin{split}
rRecall &= \frac{\sum_{i=0}^{N_A}p(L_i=1 | A_i=1, B_i=1)}{\sum_{i=0}^{N_A}p(L_i=1 | A_i=1)} * \\
& \frac{\sum_{i=0}^{N_B}p(L_i=1 | B_i=1)}{\sum_{i=0}^{N_B}p(L_i=1 | B_i=1, A_i=1)},
\end{split}
\end{equation}
where $p(L_i=1) = 1 - p(L_i=0)$ is the probability of TP for a given utterance $i$.

\subsection{\textbf{Threshold selection}}

Another important aspect to consider during a KWS model evaluation is the model sensitivity. In order to determine if a given utterance will be considered an accept or reject by the model, assuming high model scores $s_i$ are given to higher chance of detection, the utterance $i$ will be considered an accept by the model $X$, i.e., $X_i=1$, if $s_i > t_X$, where $t_X$ is a threshold attributed to model $X$ sensitivity.

Notice that $t_X$ directly impacts the rFPR and rRecall. Essentially, as we increase $t_X$, it will make the model more restrictive, so both FPR and Recall are reduced. Appendix \ref{appendix:threshold} gives an example where the threshold $t_B$ of the candidate model $B$ is found according to the trade-off between rFPR and rRecall. 

\section{Experiments}\label{sec:experiments}

In this section we present experiments to show the performance of AB/BA and ssAB/BA on simulation and real data.

\subsection{Simulations}

This section presents the simulations performed.

\subsubsection{AB/BA analysis simulation}\label{sec:abba_sim}

We created a simulation to show how the calculations from the AB/BA formulas (Equations \eqref{eq:recall_ratio} and \eqref{eq:fa_ratio}) are equivalent to the direct computation of rFPR and rRecall, but without the need for data not accepted by the models. We also show confidence intervals on those metrics as a function of the number of labels. Assume we have a source that emits positive utterances with a probability of $0.3$. The model $A$ has a Recall of $0.8$ and FPR of $0.1$. We consider two pairs of values for model $B$: Recall and FPR of ($0.82$, $0.075$) leading to rRecall and rFPR of ($1.025$, $0.75$), and Recall and FPR of ($0.84$, $0.05$) leading to rRecall and rFPR of ($1.05$, $0.5$). In addition, data accepted by model $A$ has a probability of being accepted by $B$ of $0.95$ and $0.5$ for TPs and FPs respectively. We run the simulation considering half of the data is collected by $A$, half by $B$. The simulation assumes that the model that does not collect the data is run only on the accepts of the model that does. We report estimated rFPR, rRecall in 3 scenarios, using AB/BA direct estimation, AB/BA with the introduced assumption Equation \eqref{eq:approx_method}, and using a classic A/B test (only estimating rFPR in this case). 

\begin{table*}[]
\centering
\small
\begin{tabular}{|l|l|l|l|l|l|}
\hline
\textbf{Streams} & \textbf{Labeled} & \begin{tabular}[c]{@{}l@{}}\textbf{Expected rRecall}\\ \textbf{/ rFPR Improv.}\end{tabular} & \begin{tabular}[c]{@{}l@{}}\textbf{rRecall / rFPR}\\ \textbf{Direct AB/BA}\end{tabular}                      & \begin{tabular}[c]{@{}l@{}}\textbf{rRecall / rFPR}\\ \textbf{Approx. AB/BA}\end{tabular} & \textbf{rFPR AB-Test}                     \\
\hline
10K     & 500     & 1.025 / 0.75           & \begin{tabular}[c]{@{}l@{}}1.025 {[}0.963, 1.103{]} /\\ 0.75 {[}0.48, 1.16{]}\end{tabular} & \begin{tabular}[c]{@{}l@{}}1.025 {[}0.964, 1.102{]} /\\ 0.74 {[}0.52, 1.02{]}\end{tabular} & 0.74 {[}0.43, 1.19{]} \\
\hline
100K    & 5K      & 1.05 / 0.5           & \begin{tabular}[c]{@{}l@{}}1.051 {[}1.028, 1.075{]} /\\ 0.5 {[}0.45, 0.55{]}\end{tabular}  & \begin{tabular}[c]{@{}l@{}}1.051 {[}1.028, 1.075{]} /\\ 0.49 {[}0.45, 0.54{]}\end{tabular} & 0.49 {[}0.41, 0.58{]}   \\
\hline
\end{tabular}
\caption{\textbf{Simulation of AB/BA analysis}}\label{tab:simulation}
\end{table*}

Table \ref{tab:simulation} shows the rRecall and rFPR, along with 95\% confidence interval from $1000$ bootstrapping replicates. We can notice in the table that we can detect improvement as small as a $5\%$ Recall improvement and $50\%$ FPR reduction by labeling less than $5000$ utterances. We can also see that using the assumption of same TPs and FPs between the models, represented as Approx. AB/BA in the table, to estimate $25\%$ rFPR improvement, the confidence intervals shrink from $-25\% (-52\%, +16\%)$ to $-26\% (-48\%, +2\%)$, while keeping median estimation equally accurate, showing that this is a helpful assumption. The table also shows that the rFPR predicted by regular A/B-Test on the model accepted data also gives close estimation according to the simulated parameters. However, this approach leads to higher confidence interval than the proposed approach and cannot estimate the rRecall.

\subsubsection{Semi-Supervised AB/BA analysis simulation}

One important point when working with ssAB/BA is with respect to the quality of the label generation process. Therefore, we start with a simulation showing this effect.

In our experiment, we do a Monte Carlo simulation by generating data that can be used to compute the rRecall and rFPR metrics. Data is generated such that models $A$ and $B$ collect, respectively, $40\%$ and $20\%$ as TPs. The probability that FPs and TPs from model $A$ are also accepted by model $B$ are, respectively, $0.3$ and $0.9$, and the probability that FPs and TPs from model $B$ are also accepted by model $A$ are, respectively, $0.6$, and $0.8$. Then, we simulate three soft label machines $\mathcal{M}_1$, $\mathcal{M}_2$, and $\mathcal{M}_3$ using a Beta distribution. 

Among the three machines, $\mathcal{M}_1$ is simulated to generate soft-labels with $\mathcal{B}(2,1000|L=0)$ and $\mathcal{B}(300,5|L=1)$ to have the same accuracy for both models $A$ and $B$ that will be evaluated in ssAB/BA. Then, $\mathcal{M}_2$ is simulated with $\mathcal{B}(5,100|B=1,L=0)$ and $\mathcal{B}(300,5|B=1,L=1)$ to make more mistakes in the form of higher TP probability on the FPs collected by model $B$, and $\mathcal{M}_3$ with $\mathcal{B}(2,1000|B=1,L=0)$ and $\mathcal{B}(100,10|B=1,L=1)$ to make more mistakes in the form of lower TP probability on the TPs collected by the model $B$. Notice that we keep the accuracy of the machines on model $A$ data constant, since $B$ is a candidate model with new data never seen before.

Notice that based on the parameters chosen the expected AB/BA rRecall for the simulation is $1.12$, since $P(B=1|A=1,L=1) / P(A=1|B=1,L=1) = 0.9 / 0.8 = 1.12$. Similarly, the expected rFPR is $0.5$. Results of the simulation, along with $95\%$ confidence interval from $1000$ bootstrapping replicates, are shown in Table \ref{tab:ssabba_simulation}.

\begin{table}[h]
\small
\begin{tabular}{|l|l|l|}
\hline
           & \textbf{rRecall}        & \textbf{rFPR}           \\
\hline
\textbf{AB/BA}      & 1.12{[}1.10,1.14{]} & 0.50{[}0.48,0.52{]} \\
\hline
\textbf{ssAB/BA $\mathcal{M}_1$} & 1.12{[}1.10,1.13{]} & 0.51{[}0.49,0.53{]} \\
\hline
\textbf{ssAB/BA $\mathcal{M}_2$} & 1.08{[}1.05,1.10{]} & 0.51{[}0.49,0.53{]} \\
\hline
\textbf{ssAB/BA $\mathcal{M}_3$} & 1.12{[}1.10,1.15{]} & 0.56{[}0.54,0.57{]} \\
\hline
\end{tabular}
\caption{\textbf{Simulation comparing AB/BA and ssAB/BA analysis according to label quality}}\label{tab:ssabba_simulation}
\end{table}

From Table \ref{tab:ssabba_simulation}, as expected, $\mathcal{M}_1$ results are almost exactly the same as in AB/BA, since $\mathcal{M}_1$ mean TP probabilities are $0.2\%$ for $L=0$ and $98.4\%$ for $L=1$, which are close to ground-truth. When using $\mathcal{M}_2$, however, we can see that the rRecall measured drops from $1.12$ to $1.08$, as this model makes more mistakes on the FPs collected from model $B$, increasing the machine TP probability on this data from $0.2\%$ to $4.8\%$. In this case, we can see that the rFPR is unchanged at the reported precision. Similarly, in the case of $\mathcal{M}_3$ we see a change in the reported rFPR, which increases from $0.51$ to $0.56$ as this model makes more mistakes on the TPs collected from model $B$, dropping the TP probability on this data from $98.4\%$ to $90.9\%$. In this case, the rRecall is mostly unchanged. This results show that, although the label-generation process by machine can be imperfect, it gives good approximations on the Recall Ratio and FPR Ratio in order to make deployment decisions.

\subsection{AB/BA Analysis on Real Application}

Next, we show how AB/BA performs when applied to real customer data in order to guide the decision on how much customer experience is being improved with the deployment of new KWS models. Comparison between AB/BA and ssAB/BA analysis on real data is show in Appendix \ref{appendix:ssabba_real_data}.

\subsubsection{Comparison between deployments with different threshold $t_b$}

We show results from two real deployments, called $D_1$ and $D_2$. In $D_1$, the AB/BA analysis ratio metrics are used to compare a baseline model to a candidate model with high threshold (more restrictive), while in $D_2$ the same baseline is compared to the same candidate model with low threshold (more permissive). The two deployments use about $5000$ annotated utterances per model. Results are show in Table \ref{tab:abba_deployments_op_change}.

\begin{table}[b]
\small
\begin{tabular}{|l|l|l|}
\hline
\textbf{D} & \textbf{rRecall}         & \textbf{rFPR}              \\
\hline
$D_1$          & 0.95 {[}0.94-0.96{]} & 0.57 {[}0.46-0.68{]}   \\
\hline
$D_2$          & 1.07 {[}1.05-1.09{]} & 1.33 {[}1.17 - 1.44{]}\\
\hline
\end{tabular}
\caption{\textbf{AB/BA Analysis results when deploying models with different thresholds}}\label{tab:abba_deployments_op_change}
\end{table}

As we can see in Table \ref{tab:abba_deployments_op_change}, $D_1$ resulted in loss of Recall by $5\%$ relative, but improving the rFPR in $43\%$ relative. By deploying the candidate model with low $t_B$ threshold ($D_2$), the AB/BA analysis then shows $7\% (5\%, 9\%)$ relative improvement in Recall, with a trade-off of $33\% (17\%, 44\%)$ relative increase in FPR. AB/BA analysis correctly found that the $D_1$ model is a more conservative model, and $D_2$ a more sensitive model than the baseline model.

\subsubsection{Comparison between AB/BA analysis and the evaluation a test dataset}

New candidate models are evaluated on offline test datasets to decide if they can be deployed to customers. However, offline test datasets are composed of utterances collected by the current or older deployed models, and recall improvement of those candidate models may be under-estimated. Here we show measurements from offline evaluation and AB/BA analysis in a real deployment. The AB/BA analysis was performed with approximately $7000$ annotated utterances per model.

\begin{table}[]
\small
\begin{tabular}{|l|l|l|}
\hline
\textbf{Dataset}  & \textbf{rRecall} & \textbf{rFPR}        \\
\hline
Test set & 1.03 & 0.5               \\
\hline
AB/BA    & 1.16{[}1.15-1.19{]} & 0.21{[}0.14-0.26{]} \\
\hline
\end{tabular}
\caption{\textbf{AB/BA Analysis Recall Ratio comparison to test set metrics}}\label{tab:abba_deployments_canonical_eval}
\end{table}

As we can see in Table \ref{tab:abba_deployments_canonical_eval}, the rRecall estimation from the offline test set resulted in $3\%$ relative improvement at the same FPR. That was significantly less than the $16\%$ relative improvement measured during AB/BA, in which data from model $B$ is used in the analysis. Similar observation can be made in terms of rFPR, where evaluation on the test set resulted in a $50\%$ relative improvement at the same Recall, but by also accounting for data collected by the $B$ model in AB/BA, we see that the improvements was $79\%$ relative. This shows the importance of techniques such as AB/BA analysis to better assess the customer experience on the model being deployed.

\section{Conclusion}\label{sec:conclusion}

In this paper, we presented a new framework called AB/BA analysis for recall improvement estimation of KWS under high privacy settings. We have shown that by running a candidate model offline on data collected by a baseline, and the baseline model offline on data collected by the candidate model, we were able to compute the relative Recall and relative FPR ratios using only utterances accepted by both models and use it to indicate if a candidate model is better than the baseline. We have also shown that reasonable assumptions can be used to construct an estimator with low variance, even when the number of FPs is small. Finally, we saw that a Semi-Supervised formulation of AB/BA can be used with machine-generated labels representing the probability of a true accept. This techniques brings further improvements to AB/BA analysis, especially regarding privacy on the audio collected, which does not need to be listened to and annotated.

In the past few years, much improvement has been made on the KWS systems. Evaluation metrics, however, are typically based on previously collected data from similar KWS models or data from other sources, resulting in evaluations that do not necessarily translate to customer experience, as it fails to show improvements in data never collected due to privacy constraints. Given that not much attention has been paid to this research topic, we believe that AB/BA analysis is a valuable contribution, and we hope it helps bringing interest in this line of research.

\bibliography{custom}
\bibliographystyle{acl_natbib}
\clearpage
\newpage
\appendix

\section{Appendix}\label{sec:appendix}

\subsection{Basic Evaluation Concepts}\label{sec:metrics}

During our discussion we assume that we have a classification model, $\mathcal{M}(x) \rightarrow y$, which gets an arbitrary input $x$ and output $y \in \{0, 1\}$, where $1$ and $0$ represent, respectively, a positive and negative label. 

\subsubsection{\textbf{Precision and Recall}}

The Precision and Recall metrics help to distinguish between Type-I and Type-II errors. They are defined as the following:

\begin{equation}
Precision = \frac{TP}{TP+FP}
\end{equation}

\begin{equation}
Recall = \frac{TP}{TP + FN} = \frac{TP}{P}
\end{equation}

Precision measures the proportion of correct positive predictions with respect to everything the model believes is a positive, where lower precision represents more Type-I errors. On the other hand, Recall measures the proportion of correct positive predictions with respect to all the data that is actually positive, where lower recall represents more Type-II errors. 

As illustration, in a case where 90\% of the data has label 1 and the model always gives output 1, this model will have 100\% recall, and 90\% precision. 

\subsubsection{\textbf{False Positive Rate and False Discovery Rate}}

There are also multiple ways to evaluate a model with respect to the number of False Positives (FPs) it makes. Two of them, which are explored in this paper, are False Positive Rate (FPR) and False Discovery Rate (FDR). They are defined as:

\begin{equation}
FPR = \frac{FP}{N}
\end{equation}

\begin{equation}
FDR = 1 - Precision = \frac{FP}{TP+FP}
\end{equation}

Therefore, we can see that the FPR is similar to Recall in the sense that the reference is only one class of the data, which in this case are the negative samples. It measures the proportion of negative samples that are miss classified by a model. The FDR, however, is the complement to the Precision, measuring the number of false positives among all samples that the model classifies as positive. 

Considering again the example where 90\% of the data has label 1 and the model always gives output 1, its FPR is 100\%, while its FDR is 10\%. 

\subsubsection{A/B Test}

Another related concept to our proposed method is the A/B test. The A/B test is a frequently used technique in multiple areas, such as medicine \cite{stolberg2006inventing}, marketing, political campaigning, product pricing, among others. The technique consists of doing a hypothesis test by giving two randomized and unbiased set of populations, called $A$ and $B$, two different versions of the subject being compared. For example, in the context of marketing, one could choose to give two different versions of user interface to users in order to measure differences in engagement, which is measured through statistical tests.

In the context of model evaluation, the A/B test can be explored by given two different models to users. After data collection, metrics can be computed for each population, such as metrics presented in Section \ref{sec:metrics}, and statistical tests, such as $t$-test, can be used to compare the metrics in the different populations.

The A/B test has, however limitations, when used to evaluate keyword spotting with audio privacy settings. Given that only data where the keyword is detected is collected by the models, the amount of negative data is highly biased towards the false positives from the models that collected the data. Therefore, metrics that rely on negative data, such as Recall and FPR, cannot be computed and compared using A/B test. This limitation is explored in this paper with our proposed AB/BA technique in order to tackle this challenge of Recall improvement estimation.

\subsection{Stratified Sampling}\label{appendix:stratified}

To decide if a model is better than the other one, we could simply annotate the utterances where the two models disagree. However, if it is also desirable to calculate absolute metrics, such as False Positive Rate (FPR), then annotation of model agreements is also needed. In order to reduce the amount of annotations for this task, we propose to use stratified sampling, with two strata, agreement and disagreement, such that different number of annotations are done per strata.

Our stratified sampling strategy uses the Neyman allocation principle. The optimal number of annotations $N_j$ for a strata $j$ can be found using:

\begin{equation}\label{eq:neyman}
N_j^{*} = \frac{N w_j\sqrt{p_j(1-p_j)}}{\sum_{i=1}^L w_i\sqrt{p_i(1-p_i)}},
\end{equation}
where $N$ is our annotation budget, $w_j$ is the proportion in each strata, and $p_j$ is the expected model $FPR$, assumed to be estimated, for example, from previously annotated data. 

The efficiency improvement of the stratified sampling strategy compared to a random sampling strategy, in terms of variance in the FPR, is:

\begin{equation}
\begin{split}
Eff &= 1 - \frac{\mathbb{V}({fpr}^{*})}{\mathbb{V}(fpr)} \\
&= 1 - \frac{\frac{1}{N}(\sum_j w_j\sqrt{p_j(1-p_j)})^2}{\frac{1}{N}p(1-p)},
\end{split}
\end{equation}

For the purpose of illustration, assume we have a 10\% disagreement between models, with FPR of 20\% in this strata, whereas the agreement strata has a FPR of 5\%, and the overall FPR of 8\%, Equation \eqref{eq:neyman} gives us, for the disagreement strata:

\begin{equation}
\begin{split}
\frac{N_j^{*}}{N} &= \frac{0.1\sqrt{0.2(1-0.2)}}{0.1\sqrt{0.2(1-0.2)} + 0.9\sqrt{0.05(1-0.05)}} \\
&\approx 17\%,
\end{split}
\end{equation}
indicating that the disagreement strata should optimally be 17\% of the annotation budget, without affecting the FPR variance. The efficiency gain of this method is:

\begin{equation}
\begin{split}
&Eff =\\
& 1 -  \frac{(0.1\sqrt{0.2(1-0.2)} + 0.9\sqrt{0.05(1-0.05)})^2}{0.08(1-0.08)}\\
&\approx 24\%,
\end{split}
\end{equation}
indicating that the annotation budget can be reduced by approximately 24\%, without affecting the overall FPR variance. 

\subsection{Threshold Selection Example}\label{appendix:threshold}

Given two models $A$ and $B$, where $A$ is a baseline and $B$ a candidate, we assume that $A$ has a known $t_A$, previously obtained according to the desired FPR and Recall trade-off. Therefore, in order to determine if the candidate model $B$ is better, we can calculate the FPR Ratio and Recall Ratio for multiple thresholds $t_B$. Next, the decision will be guided towards the goal of model $B$. Say, for example, that the goal of model $B$ is not to improve FPR, but only Recall, then we select $t_B = t$ such that when applying this threshold it results in $FPR\_Ratio = 1$. One example, illustrating this process, is given in Table \ref{tab:op}.

\begin{table}[h]
\small
\begin{tabular}{|l|l|l|}
\hline
\textbf{$t_B$} & \textbf{FPR Ratio} & \textbf{Recall Ratio} \\
\hline
0.1   & 1.5      & 1.20         \\
\hline
0.2   & 1.0      & 1.05         \\
\hline
0.3   & 0.8      & 1.01         \\
\hline
0.4   & 0.7      & 0.98         \\
\hline
\end{tabular}
\caption{\textbf{Threshold selection:} The table shows an example of how FPR Ratio and Recall Ratio change, as a function of $t_B$. It shows that, by selecting $t_B=0.2$, model B has the same FPR Ratio as model $B$, but improves Recall in 5\%.}\label{tab:op}
\end{table}

\begin{figure}[h]
\includegraphics[width=0.7\columnwidth]{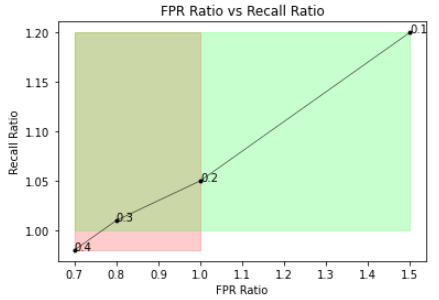}
\caption{\textbf{Model sensitivity selection:} The green area in the figure shows the region where the recall is improved, and the red area where the FPR improves. We can see that 0.2 and 0.3 are potential thresholds for $t_B$ to improve the performance of model $A$.}\label{fig:fpr_recall}
\end{figure}

Table \ref{tab:op} shows the FPR Ratio and Recall Ratio as a function of $t_B$, similar to the data behind a traditional DET curve, but using the proposed ratio metrics. We can see that, by selecting $t_B=0.2$, model $B$ has the same FPR Ratio as model $B$, but improves Recall in 5\% relative. It is also interesting to see that using $t_B = 0.3$ causes improvement in both FPR Ratio and Recall Ratio. Using $t_B=0.1$ has 20\% relative improvement in Recall, but with a high trade-off in FPR and, similarly $t_B=0.4$ has 30\% relative improvement in FPR, but with degradation in Recall. 

We see, therefore, that the model threshold can be selected according to the goal in the trade-off between FPR Ratio and Recall Ratio. It is clear though that the model $B$ is superior than model $A$, since it has a threshold region that improves both FPR and Recall, as shown in Figure \ref{fig:fpr_recall}.

\subsection{Soft-Label Score Calibration}\label{appendix:calibration}

\begin{figure}[h]
\includegraphics[width=0.7\columnwidth]{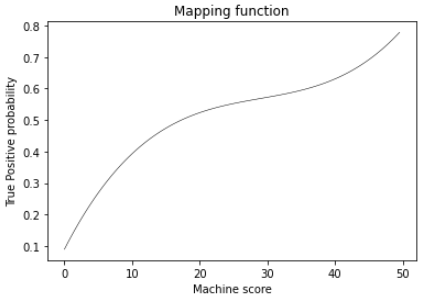}
\caption{\textbf{Probability mapping function:} Using a polynomial of degree 3, arbitrary for this illustration, we can see that scores from a model between 0 and 50 are mapped to a probability of True Positive.}\label{fig:calibration}
\end{figure}

Given a set of utterances composed of $N$ machine-generated soft labels $m \in \mathbb{R}^N$ and human labels $y \in \mathbb{R}^N$, we propose to learn $\phi(m_i)$ using a polynomial of degree 3. Notice that, as the target variable is binary, and we expect to have more TAs as $m$ increases, the polynomial should be a monotonic increasing function between $[0,1]$.  Although not guaranteed, our experiments show this to be an empirically good choice. One example is shown in Figure \ref{fig:calibration}. However, since the polynomial is not guaranteed to have probabilities bounded between $[0,1]$, we have also explored to use a $B$-Spline Logistic Regression with monotonic constraints \citep{eilers1996flexible, barlow1972isotonic}. However we have not seen significant difference and have decided to use the polynomial approach for simplicity.

Once we have soft-labels for the utterances from model $A$ and $B$ to be compared, we can use Equations \eqref{eq:fa_ratio_ssl} and \eqref{eq:recall_ratio_ssl} in order to estimate the FPR Ratio and Recall Ratio trade-off. 

\subsection{Semi-Supervised AB/BA Comparison to AB/BA on Real Data}\label{appendix:ssabba_real_data}

Next we show real data examples where we used both AB/BA and ssAB/BA, in order to see if the ratios reported are similar. In this case, machine-generated scores were generated by a cloud-side verification system, and its scores converted to a TP probability according to a polynomial mapping function learned from other existing labeled datasets, following the process described in Section \ref{sec:ssabba}. Results are shown in Table \ref{tab:ssabba_real}.

\begin{table}[h]
\small
\begin{tabular}{|l|l|l|}
\hline
\textbf{Example} & \begin{tabular}[c]{@{}l@{}}\textbf{ssAB/BA rRecall}\\ \textbf{at rFPR}\end{tabular} & \begin{tabular}[c]{@{}l@{}}\textbf{AB/BA rFPR}\\ \textbf{at rRecall}\end{tabular} \\
\hline
1       & 1.03 {[}1.03-1.03{]}                                                          & 1.04{[}1.04-1.05{]}                                                           \\
\hline
2       & 1.01 {[}1.01-1.01{]}                                                          & 1.0{[}0.97-1.05{]}                                                            \\
\hline
3       & 0.99 {[}0.98-1.01{]}                                                          & 1.01{[}0.99-1.04{]}\\
\hline
\end{tabular}
\caption{\textbf{Semi-Supervised AB/BA and AB/BA analysis comparison on real data:} The table shows three examples comparing AB/BA and Semi-Supervised AB/BA on real customer data}\label{tab:ssabba_real}
\end{table}

Results in Table \ref{tab:ssabba_real} show that ssAB/BA is able to well-approximate the AB/BA results, having results with overlapping margin of error in most cases. It is important to notice, however, that there is a risk of using ssAB/BA related to the quality of labels generated. In the Example 3, we can see that ssAB/BA results suggest a Recall loss of $1\%$ relative, while AB/BA suggests a Recall improvement of $1\%$ relative, although the confidence intervals overlap. That represents the case where, when the recall improvement is small, the uncertainty of the machine-label generation may limit its applicability. It is important, therefore, to monitor the quality of the label machines in order to know how trustworthy they are, and to also use other auxiliary metrics that help reducing the risk of trusting the ssAB/BA results by itself.

\end{document}